\documentclass{PoS}

\title{The Phases of Non-Compact QED$_3$}

\ShortTitle{The Phases of Non-Compact QED$_3$}

\author{\speaker{Costas Strouthos}\\
         Institute of Chemical Sciences and Engineering, \'Ecole Polytechnique F\'ed\'erale de Lausanne, \\
         1015 Lausanne, Switzerland \\
         E-mail: \email{cstrouth@deas.harvard.edu}}

\author{John B. Kogut \\
        Department of Energy, Division of High Energy Physics, Washington, DC 20575 USA \\
        and \\
        Dept. of Physics - TQHN, Univ. of Maryland, 82 Regents Dr., College Park, MD 20742, USA \\
        E-mail: \email{jbkogut@umd.edu}}

\abstract{
Non-compact three-dimensional QED is studied by computer simulations
to understand its chiral symmetry breaking features for different values 
of the number of fermion flavors $N_f$.
We consider the four-component formulation for the fermion fields,
which arises naturally as the continuum limit of the staggered fermion
construction in $(2+1)$ dimensions. We present preliminary results for
the equation of state of the theory in an effort to understand the properties
of the chiral phase transition of the theory at a critical number of fermion flavors $N_{fc}$. 
Our preliminary results indicate that $N_{fc} \approx 1.5$.
}

\FullConference{The XXV International Symposium on Lattice Field Theory\\
		 July 30-4 August 2007\\
		 Regensburg, Germany}

\begin{document}

\section{Introduction}

The study of QFTs in which the ground state shows a sensitivity to the number of fermion flavors $N_f$ is
intrinsically interesting. According to approximate solutions of continuum Schwinger-Dyson equations (SDEs), 
QED$_3$ displays this phenomenon.
It is believed to be confining and exhibit features such as dynamical mass
generation when the number of fermion flavors $N_f$ is smaller than a critical
value $N_{fc}$ \cite{pisarski, miranskii, pennington, maris96, nash, tesanovic.03, maris05}.
Apparently, for $N_f > N_{fc}$, the attactive interaction between a fermion and
an antifermion due to photon exchange is overwhelmed by the fermion
screening of the theory's electric charge.
Furthermore, over the past few years QED$_3$ has attracted a lot of attention, because of
potential applications to models of high $T_c$ superconductivity 
\cite{dorey, tesanovic.02, tesanovic.03, herbut, mavromatos}
It is also an interesting  and challenging model field theory which is
being seen as an ideal laboratory to study more complicated gauge field theories.

Initial studies based on SDEs using the
photon propagator derived from the leading order $1/N_f$ expansion, where $N_f$
is the number of fermion flavors, suggested
that for $N_f$ less than $N_{fc} \simeq3.2$
chiral symmetry is broken \cite{pisarski}. The model in the limit
$N_f\to N_{fc}$ is supposed to undergo an infinite-order phase transition \cite{miranskii}. 
Other studies
taking non-trivial vertex corrections into account predicted chiral symmetry
breaking for arbitrary $N_f$ \cite{pennington}. Studies which treat
the vertex consistently in both numerator and denominator of the SDEs
have found $N_{fc}<\infty$, with a value either in agreement with the
original study \cite{maris96},
or slightly higher $N_{fc}\simeq4.3$ \cite{nash}.
An argument based on a thermodynamic inequality  
predicted $N_{fc}\leq{3\over2}$ \cite{appelquist}, a result that was later on challenged  
in \cite{mavromatos}. A gauge invariant determination of $N_{fc}$ based on the 
divergence of the chiral susceptibility gives $N_{fc} \approx 2.16$ \cite{tesanovic.03}.  
Recent progress in the direction of gauge covariant solutions for the propagators of QED$_3$ 
showed that in the Landau gauge a chiral phase transition exists 
at $N_{fc} \approx 4$ \cite{maris05}. Furthermore, a perturbative analysis of RG flow equations in the
large-$N_f$ limit predicts $N_{fc} \approx 6$ \cite{kaveh}.

There have also been numerical attempts to resolve the issue via lattice 
simulations of QED$_3$.
Once again, opinions have divided on whether $N_{fc}$ is
finite \cite{sasha,dagotto}, or whether chiral symmetry is broken for all $N_f$ \cite{azcoiti}. 
A numerical study of the quenched ($N_f=0$) case has shown that chiral symmetry is broken \cite{hands90}. 
More recent numerical results showed that chiral symmetry is also broken for
$N_f=1$, whereas $N_f=2$ appeared chirally symmetric with an upper bound of $10^{-4}$
on the dimensionless condensate \cite{hands2002}.
The principal obstruction to a definitive answer has been
large finite volume effects resulting from the presence of a massless photon in
the spectrum, which prevent a reliable extrapolation to the thermodynamic limit.
Recent lattice simulations of the three-dimensional Thirring model, which may have the same 
universal properties as QED$_3$, predicted $N_{fc}=6.6(1)$ \cite{thirring}.  
In this paper we present a study of the QED$_3$ Equation of State based on preliminary results 
extracted from lattice simulations on large lattices (up to $80^3$). 

\section{The Model}
We are considering the four-component formulation of QED$_3$ where the Dirac
algebra is represented by the $4 \times 4$ matrices $\gamma_0$, $\gamma_1$
and $\gamma_2$. This formulation preserves parity and gives each spinor
a global $U(2)$ symmetry generated by
$\bf1$, $\gamma_3, \gamma_5$ and $i\gamma_3 \gamma_5$; the full
symmetry is then $U(2N_f)$. If the fermions acquire dynamical mass
the $U(2N_f)$ symmetry is broken spontaneously to $U(N_f) \times U(N_f)$
and $2N_f^2$ Goldstone bosons appear in the particle spectrum.

The action of the lattice model we study is
\begin{eqnarray}
S &=&\frac{\beta}{2} \sum_{x,\mu<\nu} \Theta_{\mu \nu}(x) \Theta_{\mu \nu}(x)
+ \sum_{i=1}^N \sum_{x,x^\prime} {\bar \chi}_i(x) M(x,x^\prime)
\chi_i(x^\prime)
\label{eq:action}\\
\Theta_{\mu \nu}(x) &\equiv& \theta_{x\mu}+\theta_{x+\hat\mu,\nu}
-\theta_{x+\hat\nu,\mu}-\theta_{x\nu}\nonumber\\
M(x,x^\prime) &\equiv&
m\delta_{x,x^\prime}+\frac{1}{2} \sum_\mu\eta_{\mu}(x)
[\delta_{x^\prime,x+\hat \mu} U_{x\mu}
-\delta_{x^\prime,x-\hat \mu} U_{x-\hat \mu,\mu}^\dagger].\nonumber
\end{eqnarray}
This describes interactions between $N$ flavors of
Grassmann-valued staggered fermion fields
$\chi,\bar\chi$ defined on the sites $x$ of a three-dimensional cubic lattice, and
real photon fields $\theta_{x\mu}$ defined on the link between nearest neighbour
sites $x$, $x+\hat\mu$. Since $\Theta^2$ is unbounded from above, eq.(\ref{eq:action})
defines a
non-compact formulation of QED; note however that to ensure local gauge
invariance the fermion-photon interaction is encoded via the compact connection
$U_{x\mu}\equiv\exp(i\theta_{x\mu})$, with $U_{x+\hat\mu,-\mu}=U^*_{x\mu}$.
In the fermion kinetic matrix $M$ the Kawamoto-Smit phases
\begin{equation}
\eta_\mu(x)=(-1)^{x_1+\cdots+x_{\mu-1}}
\label{eq:KS}
\end{equation}
are designed to
ensure relativistic covariance in the continuum limit, and $m$ is the bare
fermion mass.

If the physical lattice spacing is denoted $a$, then
in the continuum limit $a\partial\to0$, eq.(\ref{eq:action}) can be shown to be
equivalent up to terms of $O(a^2)$ to
\begin{equation}
S=\sum_{j=1}^{N_f}\bar\psi^j[\gamma_\mu(\partial_\mu+igA_\mu)+m]\psi^j
+{1\over4}F_{\mu\nu}F_{\mu\nu}
\label{eq:Scont}
\end{equation}
ie. to continuum QED in 2+1 euclidean dimensions,
with $\psi,\bar\psi$ describing $N_f$ flavors of four-component Dirac spinor acted
on by 4$\times$4 matrices $\gamma_\mu$, and
$N_f\equiv2N$. The continuum photon field is related to the lattice field via
$\theta_{x\mu}=agA_\mu(x)$, with dimensional coupling strength $g$ given
by $g^2=(a\beta)^{-1}$, and the field strength
$F_{\mu\nu}=\partial_\mu A_\nu-\partial_\mu A_\nu$. The continuum limit is thus
taken when the dimensionless inverse coupling $\beta\to\infty$.
                                                                                                                      
As reviewed in \cite{hands2002}, for $a>0$ in the chiral limit
the lattice action (\ref{eq:action}) retains only a remnant of the
U(2$N_f$) global symmetry of (\ref{eq:Scont})
under global chiral/flavor rotations, namely a
U($N)\otimes\mbox{U}(N)$ symmetry which is broken to U($N$) either explicitly by
the bare mass $m\not=0$, or spontaneously by a chiral condensate
$\langle\bar\chi\chi\rangle\not=0$, in which case the spectrum contains
$N^2$ exact Goldstone modes. It is expected that the symmetry breaking
pattern  $\mbox{U}(2N_f)\to\mbox{U}(N_f)\otimes\mbox{U}(N_f)$ is restored
in the continuum limit, implying the existence of an additional 7$N^2$
approximate Goldstone modes whose masses vanish as $\beta\to\infty$.

\section{Numerical Simulations}

\begin{figure}
\begin{center}
\includegraphics[width=.7\textwidth]{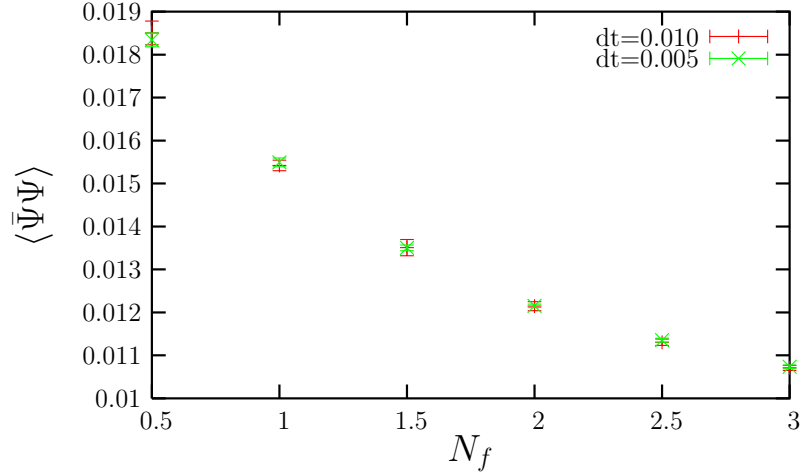}
\caption{Chiral condensate versus $N_f$ from simulations with $dt=0.010$ and $dt=0.005$; $\beta=0.90$, $m=0.005$, and $L=80$.}
\label{fig:fig1}
\end{center}
\end{figure}

In this section we present results from numerical simulations performed using the 
standard Hybrid Molecular Dynamics (HMD) R-algorithm. In order to ensure that the $O(N^2 dt^2)$ 
systematic errors ($dt$ is the time step of the HMD trajectory) 
are negligible, we performed two sets of simulations one with  $dt=0.010$ and another with $dt=0.005$ 
on $80^3$ lattices with $m=0.005$ and $\beta = 0.90$. 
As shown in Fig.~\ref{fig:fig1} the values of the chiral condensate from the two sets 
of simulations with $N_f=0.5, ..., 3.0$ agree within statistical error, implying that 
the systematic effects are negligible.

\begin{figure}
\begin{center}
\includegraphics[width=.7\textwidth]{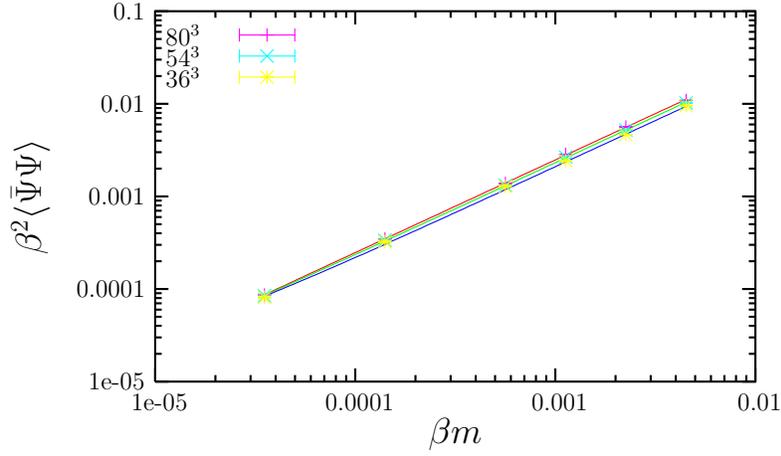}
\caption{Dimensionless condensate versus dimensionless bare mass for $N_f=1.5$.}
\label{fig:fig2}
\end{center}
\end{figure}

\begin{figure}
\begin{center}
\includegraphics[width=.7\textwidth]{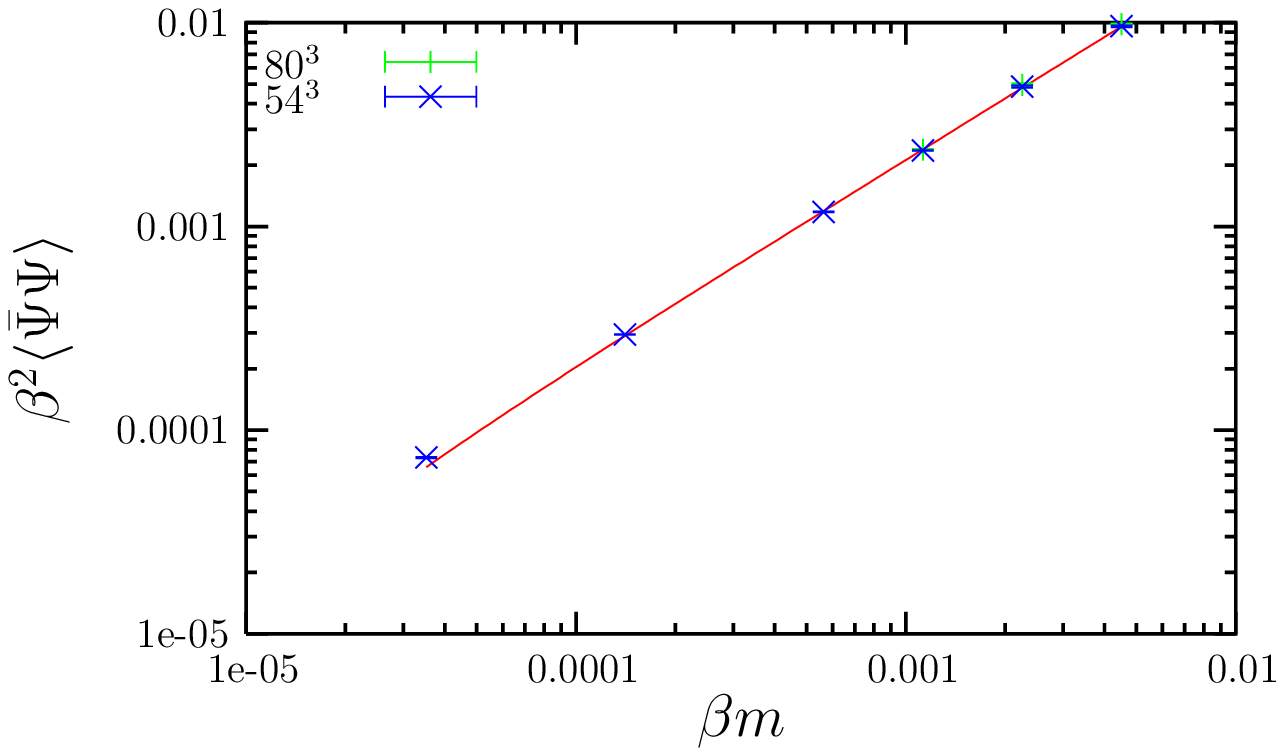}
\caption{Dimensionless condensate versus dimensionless bare mass for $N_f=2.0$.}
\label{fig:fig3}
\end{center}
\end{figure}

Next we present results from simulations with $N_f=1.5$ and $N_f=2.0$ 
(Figs.~\ref{fig:fig2} and \ref{fig:fig3}). More specifically, we study the behavior 
of the dimensionless chiral condensate $\beta^2 \langle \bar{\psi} \psi \rangle$ versus   
the dimensionless fermion bare mass $\beta m$ on different lattice sizes. In the $N_f=1.5$
case the finite size discrepancy between the $L=54$ and $L=80$ results is $\sim 7\%$, implying that 
the finite size effects for $L=80$ are small. In the $N_f=2.0$ case the discrepancy is $\sim 1\%$, 
implying that in this case the values of the condensate extracted from 
the $80^3$ simulations should be even closer to the thermodynamic limit values. 

For $N_f=2$, we fitted the $54^3$ data to 
\begin{equation}
\beta^2 \langle \bar{\psi} \psi \rangle = a_0 + a_1\cdot(\beta m) + a_2\cdot(\beta m)^2.
\label{eq:fit1}
\end{equation}
The extracted value of  $a_0 = -8 \times 10^{-8}$ with a statistical error $5 \times 10^{-7}$ and fit quality 
$\chi^2/DOF=0.9$ is consistent with zero with relatively high accuracy. 
For $N_f=1.5$, we fitted the $80^3$ data to (\ref{eq:fit1}) and got $a_0 = -2 \times 10^{-6}$ with a statistical error 
$10^{-6}$ and fit quality $\chi^2/DOF= 4.1$. This result is also very close to zero although the 
fit quality is lower than in the $N_f=2.0$ case. As we will se below the low fit quality could be attributed to 
the fact that $N_f=1.5$ may be close to the chiral phase transition. 

\begin{figure}
\begin{center}
\includegraphics[width=.7\textwidth]{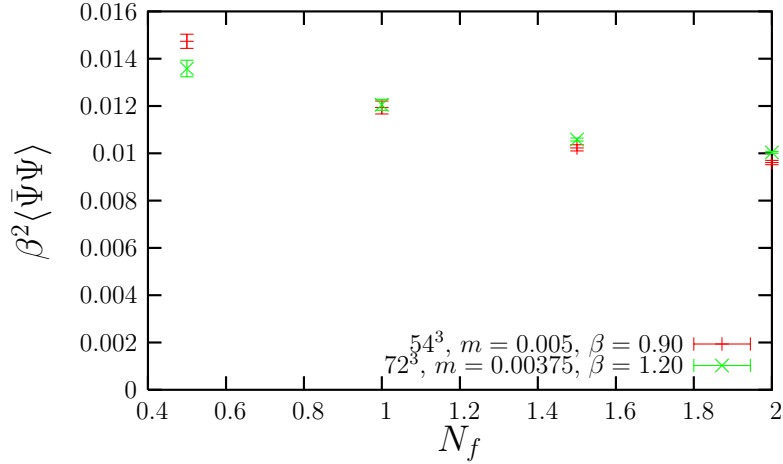}
\caption{Dimensionless condensate versus $N_f$ at $\beta=0.90$ and $1.20$ with constant $L/\beta$ and $\beta m$.}
\label{fig:fig5}
\end{center}
\end{figure}

\begin{figure}
\begin{center}
\includegraphics[width=.5\textwidth]{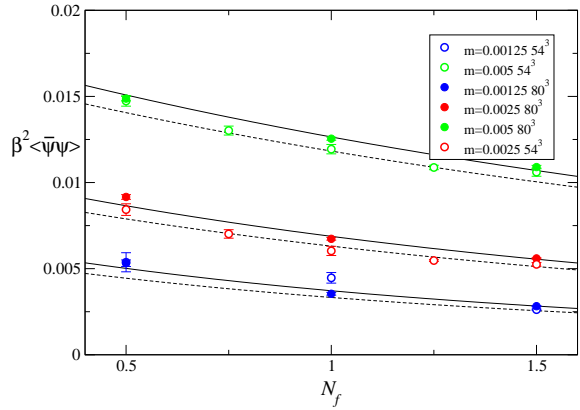}
\caption{Fits of $ \beta^2 \langle \bar{\Psi} \Psi \rangle$ vs. $N_f$ to a finite volume scaling
form of the Equation of State.}
\label{fig:fig4}
\end{center}
\end{figure}
We also checked the effects of lattice discretization on the values of the chiral condensate   
by comparing data extracted from simulations with $\beta=0.90$ and $\beta=1.20$ for $N_f=0.5,...,2.0$. 
We achieved this by fixing for the two sets of simulations the physical volume $(L/\beta)^3$ 
and the physical mass $\beta m$. The results presented in Fig.\ref{fig:fig5} show that the lattice discretization 
effects at $\beta=0.90$ are small 
for $N_f>0.5$, whereas for $N_f=0.5$ there is an $\sim 8\%$ discrepancy between the values of the condensate at 
$\beta=0.90$ and $\beta=1.20$. 

Next, we fitted the data for the dimensionless condensate at different values of $N_f$ 
and $m$, and lattice sizes $54^3$ and $80^3$ to a renormalization group inspired equation 
of state that includes a finite size scaling term \cite{deldebbio}:
\begin{equation}
m=A((\beta-\beta_c)+CL^{-{1\over\nu}}) (\beta^2 \langle\bar{\psi}\psi\rangle)^p+
B (\beta^2 \langle\bar{\psi}\psi\rangle)^\delta ,
\end{equation}
where $p=\delta-1/\beta_m$.
The results extracted from this fit are:
$A=0.0477(38),  B=0.79(2), C=10.7(8), N_{fc}=1.52(6), \delta=1.177(7), p=0.73(2)$. 
The data and the fitting functions are shown in Fig. \ref{fig:fig4}.
These results are consistent with a relatively smooth second order phase transition.

\section{Summary}
We studied numerically the equation of state of non-compact QED$_3$. The extrapolations  
to the chiral limit on lattices with small finite size effects show with high accuracy 
that the $N_f=2.0$ theory is chirally symmetric with an accuracy of $O(10^{-7})$. 
The preliminary results extracted from fits to a finite volume 
Equation of State are consistent with a second order phase transition scenario and $N_{fc} \approx 1.5$. 
However, in order to reach a decisive conclusion on the value of $N_{fc}$ 
and the properties of the chiral phase transition we have to extend our simulations to larger lattices, 
generate better statistics, and include more data points close to the continuum limit. 

\section*{Acknowledgements}
Discussions with Simon Hands and Pavlos Vranas are greatly appreciated.

\end{document}